\documentclass[aip,graphicx,reprint,apl]{revtex4-1}

\usepackage{graphicx}
\usepackage{dcolumn}
\usepackage{bm}

\usepackage[utf8]{inputenc}
\usepackage[T1]{fontenc}
\usepackage{mathptmx}
\usepackage{etoolbox}
\usepackage{siunitx}

\begin{document}


\title{Magnetic field--controlled nanoscale spin-wave vertical directional coupler} 



\author{Krzysztof Szulc}
\email{szulc@ifmpan.poznan.pl}
\affiliation{Institute of Molecular Physics, Polish Academy of Sciences, M. Smoluchowskiego 17, 60-179, Poznań, Poland}
\affiliation{Institute of Spintronics and Quantum Information, Faculty of Physics and Astronomy, Adam Mickiewicz University, Uniwersytetu Poznańskiego 2, 61-614, Poznań, Poland}

\author{Maciej Krawczyk}
\affiliation{Institute of Spintronics and Quantum Information, Faculty of Physics and Astronomy, Adam Mickiewicz University, Uniwersytetu Poznańskiego 2, 61-614, Poznań, Poland}


\date{\today}

\begin{abstract}
The directional coupler is a fundamental element of wave-based circuits. The state of the art for the spin-wave directional couplers consists mostly of macroscopic waveguides or two-dimensional planar systems. In this Letter, we present the design of the nanoscale spin-wave vertical directional coupler with a very high efficiency exceeding 99.5\%. We demonstrate that the operation of the coupler can be controlled by the magnitude of the external magnetic field. Moreover, it can perform multiplexing and demultiplexing of the spin-wave signal. Such a device ought to become an essential element of the three-dimensional magnonic circuits.
\end{abstract}


\maketitle 


Scientists are currently searching for a technology that can replace or supplement electronics in the computer market. One of the main contenders is magnonics.\cite{barman2021magnonics,flebus2024magnonic} The spin wave is a promising information carrier because it does not require the transmission of charges, creating the possibility for energy-efficient computing.\cite{mahmoud2020introduction,chumak2022advances} Computing schemes for magnonics include not only Boolean logic,\cite{khitun2005nano,khitun2010magnonic} but also analog,\cite{csaba2014spin,khivintsev2016prime} quantum,\cite{demokritov2006bose,lachance2020entanglement} neuromorphic,\cite{arai2018neural,papp2021nanoscale}, reservoir,\cite{watt2021implementing} and non-volatile computing.\cite{khitun2011non} Many devices have been already proposed including logic gates,\cite{kostylev2005spin,schneider2008realization,lee2008conceptual,rana2018voltage} majority gates,\cite{klingler2014design,fischer2017experimental,talmelli2020reconfigurable} transistors,\cite{chumak2014magnon} phase shifters,\cite{vasiliev2007spin,demidov2009control} diodes,\cite{lan2015spin,szulc2020spin,grassi2020slow} circulators,\cite{roberjot2021multifunctional,wang2021inverse} multiplexers,\cite{heussner2018frequency} and half-adders.\cite{wang2020magnonic} In all of these applications, effective signal transmission is essential, and information must be carried through and between channels. The latter can be achieved using directional couplers.

Magnonic couplers allow the signal to be transmitted between two waveguides due to the dynamic stray fields generated by spin waves. The wave generated in one waveguide acts on the magnetization in the second waveguide, resulting in the excitation of the spin wave and hence the transmission of the signal. Most of the spin-wave directional couplers are based on the planar systems which are relatively easy to fabricate,\cite{sadovnikov2015directional,sadovnikov2016nonlinear,sadovnikov2017toward,wang2018reconfigurable,ren2019reconfigurable,wang2020magnonic} while the three-dimensional systems so far are of macroscopic size.\cite{grachev2024nonreciprocal}

In this Letter, we show that the nanoscale spin-wave vertical directional couplers can be more advantageous for applications than planar ones, opening up new possibilities inaccessible in planar and macroscopic systems. Firstly, nanoscale waveguides have a wide range of frequencies with only one mode available. Secondly, the system of the vertically coupled waveguides has a critical frequency $f_{\rm c}$ at which the frequencies of the coupler modes of different symmetry interchange. Since $f_{\rm c}$ varies with the external magnetic field, we propose the design of a sub-\SI{}{\micro\meter} directional coupler controllable with the magnitude of the bias magnetic field. In addition, we show that it can also work as a multiplexer and a demultiplexer.


\begin{figure*}
    \includegraphics{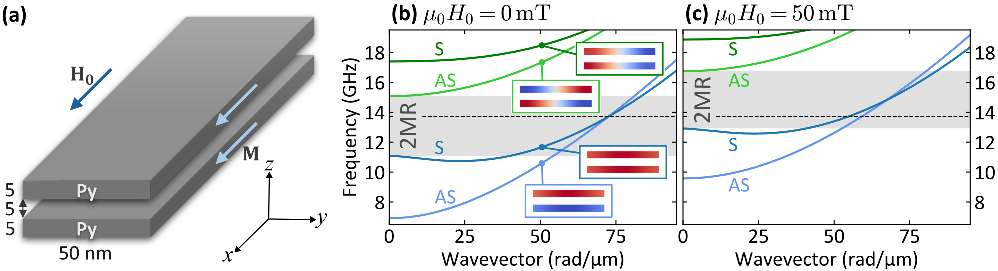}
    \caption{(a) Scheme of the investigated system of two identical Py waveguides. (b-c) The dispersion relation of the pair of infinitely long waveguides (b) for $\mu_0 H_0=0$ and (c) $\mu_0 H_0=50$~mT. The blue lines indicate the fundamental modes, while the green lines are first-order standing modes quantized over the width. S denotes the symmetric modes while AS indicates the antisymmetric modes. The gray area marks the two-mode range (2MR). In (b), the inset plots show the dynamic component of the magnetization $m_y$ at $k=\SI{50}{\radian/\micro\meter}$ in the waveguides' cross-sections, where red color indicates the positive value of $m_y$ and blue color indicates the negative value. The horizontal dashed line indicates the frequency of 13.9~GHz where the S and AS modes cross at zero field. 
    \label{fig:disp}}
\end{figure*}

The building block of our directional coupler is a waveguide made of permalloy (Py, Ni$_{80}$Fe$_{20}$) with 50~nm width and 5~nm thickness. The system consists of two stacked waveguides separated by a 5~nm non-magnetic medium, as shown in Fig.~\ref{fig:disp}(a). The waveguides are uniformly magnetized along the long axis with a parallel relative orientation. We assumed typical magnetic parameters of Py, i.e., saturation magnetization $M_{\rm s} = \SI{850}{\kilo\ampere/\meter}$, exchange constant $A_{\rm ex} = \SI{13}{\pico\joule/\meter}$, and gyromagnetic ratio $\gamma = \SI{28}{\giga\hertz/\tesla}$. The effect of the damping is neglected.

The first step in designing the directional coupler is to investigate the dispersion relations of the coupled elements. This is done by numerical simulations using COMSOL Multiphysics.\cite{vanatka2021spin} The details of the simulations are presented in Section I of the Supplementary Material.

We calculated the dispersion relation for two cases -- the absence of external magnetic field [Fig.~\ref{fig:disp}(b)] and the external field of 50~mT applied along the long axis of the waveguides [Fig.~\ref{fig:disp}(c)]. Since the spin-wave propagation direction is parallel to the magnetization direction, the dispersion branches have the character of backward volume magnetostatic modes\cite{stancilprabhakar} (although not all modes have the character of backward waves). In the investigated frequency range (up to 20~GHz), four modes are present. Due to the coupling between the waveguides, the modes form pairs of symmetric (S) and antisymmetric (AS) modes (see inset plots in Fig.~\ref{fig:disp}(b)). Two lower modes, marked in blue, are fundamental modes of the waveguides, i.e., they have no nodes in the cross-section. Two green dispersions are related to the standing modes with a first-order quantization over the width of each waveguide. For small wavevectors, the AS mode has a lower frequency, while for large wavevectors, the order is reversed. At a critical wavevector $k_{\rm c}$, the S and AS branches cross.\cite{wang2022magnon} This point indicates a situation where the selected spin-wave modes in two waveguides do not interact with each other. For the fundamental mode in Figs.~\ref{fig:disp}(b-c), $k_{\rm c} \approx \SI{70}{\radian/\micro\meter}$. This mode crossing is inherent to the systems of vertically stacked waveguides (strictly speaking, waveguides facing each other's longer side), as it results from the change in the character of the dynamic stray magnetostatic field responsible for the coupling as the wavenumber increases. In the systems of horizontally aligned waveguides, i.e. waveguides facing shorter sides, the S mode has a lower frequency for all wavevectors. The second important property of this system is a relatively wide frequency range---reaching several or even tens of GHz---of a two-mode dispersion relation. This is possible because of the small width of the waveguides -- the quantized modes have a significantly higher frequency than the fundamental modes. This feature protects the fundamental mode from being scattered into other modes, which can happen in the multimode range when translational symmetry is broken.\cite{vogt2012spin,sadovnikov2017spin}

From the perspective of the propagating waves, the coupling between the waveguides can be described by three parameters: coupling strength, coupling length $l_{\rm c}$, and transmission time.\cite{yariv1973coupled,haus1991coupled,graczyk2018co} Since the waveguides have the same size and magnetic parameters, they also have identical dispersion relations when isolated. Thus, the coupling strength for ideal infinite waveguides is always 100\%, which means that the wave can be fully transmitted between them. The coupling length is defined as the distance required to achieve the maximum possible transfer of the spin wave of a given frequency $f$ between two waveguides, and it depends on the wavevectors of the S and AS modes, $k_{\rm S}$ and $k_{\rm AS}$, respectively:
\begin{equation}
    l_{\rm c}(f) = \frac{\pi}{|k_{\rm AS}(f)-k_{\rm S}(f)|}.
\end{equation}

\begin{figure}[!b]
    \includegraphics{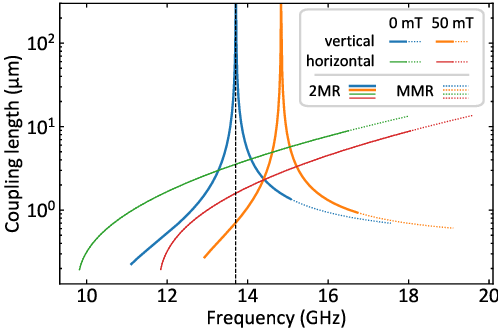}
    \caption{Coupling length as a function of the spin-wave frequency. The plot shows the comparison between the vertically stacked waveguides with a separation of 5~nm (blue and orange lines) and the planar waveguides with a separation of 25~nm (green and red lines). The blue and green lines show the dependence at 0~mT, while orange and red lines at 50~mT. The solid part of the lines represents the value in the two-mode range (2MR) while the dotted part is calculated in the multimode range (MMR). \label{fig:kappa}}
\end{figure}

Figure~\ref{fig:kappa} shows the coupling length for the pair of fundamental modes as a function of frequency for two external magnetic fields---0 (blue line) and 50~mT (orange line)---corresponding to the dispersion relations shown in Fig.~\ref{fig:disp}(b) and Fig.~\ref{fig:disp}(c), respectively. In addition, we present the coupling length for the system of horizontally coupled planar waveguides with the separation of 25~nm, where the green line is for zero magnetic field and the red line is for the field of 50~mT. Since the coupling length is inversely proportional to the difference of the wavevectors, the crossing point is represented by the parameter's maximum value approaching infinity. At zero external field, $l_{\rm c}(f)$ reaches infinity at about 13.7~GHz, which is shown by the horizontal dashed black line in Figs.~\ref{fig:disp}(b,c) and the vertical line in Fig.~\ref{fig:kappa}. Below and above this frequency, the coupling length decreases strongly, reaching the values below 1~$\SI{}{\micro\meter}$. With an increase of the external magnetic field to 50~mT, the dispersion relation goes up in frequency, and the same happens to the coupling length curve, which shifts to higher frequencies. Because of that, the coupling length at 13.7~GHz frequency decreases to about 705~nm. This makes it possible to design a directional coupler of a single coupling length whose operation (on-off) can be controlled by the external magnetic field. Furthermore, at zero external field, the same value of the coupling length (i.e., 705~nm) is also at about 12.3~GHz, a property that can be used to design the multiplexer and demultiplexer.

For the horizontally coupled waveguides, the coupling length increases monotonically with the increase of the frequency. Although the small coupling lengths are present for small frequencies (i.e., less than 1~$\SI{}{\micro\meter}$ below 11.1 and 13 GHz at 0 and 50 mT fields, respectively), they are also associated with much slower waves than in higher frequencies. Since the coupling length does not change significantly with the change in frequency, the ultralow spin-wave transmission can only be achieved when the difference between the external fields is very large. To avoid a large external field, the directional coupler must be based on the change between the odd and even number of coupling lengths with the field change,\cite{wang2018reconfigurable} thus the minimum length of the coupler will be equal to two coupling lengths. The combination of these two factors causes the directional coupler based on horizontally coupled waveguides to be significantly longer or to use much slower waves than the coupler based on the same waveguides but coupled vertically. The comparison between the transmission times for vertical and horizontal coupler, presented in Section II of the Supplementary Material, shows that the horizontal directional coupler should have longer operation time than the vertical coupler demonstrated here.

In the second stage of the study, we performed the micromagnetic simulations in Boris Computational Spintronics\cite{boris} to study the propagation of the spin waves through the directional coupler. The details of the simulations are presented in Section I of the Supplementary Material.

\begin{figure}
    \includegraphics[width=\linewidth]{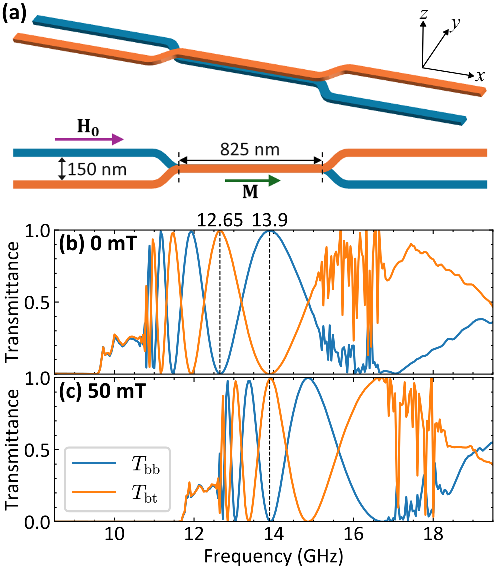}
    \caption{(a) The scheme of the vertical directional coupler in three dimensions (top) and the projection from the top with the dimensions marked (bottom). (b-c) The transmittance of the spin wave through the directional coupler (b) in zero external field and (c) in 50~mT field for the excitation of the wave on the left side of the bottom waveguide. The solid blue line represents the transmission detected in the bottom waveguide ($T_{\rm bb}$) while the solid orange line shows the transmission in the top waveguide ($T_{\rm bt}$). \label{fig:design}}
\end{figure}

The design of the directional coupler requires the proper testing of the geometry and the functionality. We have found that the system shown in Fig.~\ref{fig:design}(a) gives the best results. Each waveguide consists of a straight input and output, a straight coupler section, and two S-shaped bends. There is a number of important features that must be met to achieve the required functionality. The distance between the inputs and between the outputs should not be smaller than 100~nm to prevent the coupling between the waveguides outside of the coupler area. We chose a distance of 150~nm, which results in a coupling length of more than $\SI{100}{\micro\meter}$ in the working frequency range. The design of the S-shaped bends is based on the two interconnected arcs of the identical rings. In this geometry, the optimal angle of the arcs, for which the transmission was the highest, was 45 degrees. The size of the rings is based on the waveguide width, the distance between waveguides, and the arc angle, thus the outer radius of the ring is 195.7~nm, while the inner radius is 145.7~nm. Finally, it was found that the waveguides must cross their paths (as shown in Fig.~\ref{fig:design}(a)), otherwise a significant spin-wave intensity remains in the other waveguide. Moreover, the way of crossing is also defined. The bottom waveguide, when viewed from above, must have a pair of Z-shaped bends, while the top waveguide---a pair of S-shaped bends. The results show that the transmission is less efficient if the waveguide turns back on its path, i.e. performs opposite order of turns, e.g., first Z-shaped and second S-shaped (see e.g. Ref.~\onlinecite{wang2018reconfigurable}). The static magnetization configuration of the directional coupler is presented in Section III of the Supplementary Material.

The test of the vertical directional coupler for the selected geometry is shown in Figs.~\ref{fig:design}(b-c). They show the transmittance of the spin wave excited on the left side (as seen in Fig.~\ref{fig:design}(a)) of the bottom waveguide at zero external field (Fig.~\ref{fig:design}(b)) and 50~mT field (Fig.~\ref{fig:design}(c)). As a result of a preliminary analysis, the length of the coupler was set to 825~nm. The wave is then detected on the right side of the bottom (blue line) and top waveguide (orange line). The calculation of the transmittance was done by calculating the fast Fourier transform (FFT) in an area before and after the coupler and calculating the ratio between these values. The signal was filtered using the FFT so that only the fundamental mode is included in the transmittance calculation. 

In both fields, the transmittance fluctuates in a sinusoidal manner in the two-mode frequency range. The regular transmission starts at the minimum frequency of the S fundamental mode (10.7~GHz for 0~mT and 12.6~GHz for 50~mT) and ends at the frequency of 15.1~GHz for 0~mT or 16.6~GHz for 50~mT, where the width-quantized mode appears and the transmittance becomes chaotic. This is due to the fact that the wave is partially transferred into the quantized modes or reflected. These processes occur at the bends of the waveguides and indicate the importance of optimizing the bend areas, particularly in multimode operation.\cite{mieszczak2020anomalous,klima2024zero} Therefore, to avoid any chaotic performance, we will work only in the two-mode regime. From Figs.~\ref{fig:design}(b-c) we see that the maximum transmission to the bottom waveguide at the zero field (no coupling with the top waveguide) and to the top waveguide at 50~mT occurs for the same frequency of 13.9~GHz, which is slightly larger than predicted from the analysis of the dispersion relation. Moreover, at zero field, the full transmission to the top waveguide occurs for a few frequencies, with the largest at 12.65~GHz. This effect is to perform the multiplexing and demultiplexing of the signal with the same coupler.

\begin{figure}
    \includegraphics[width=\linewidth]{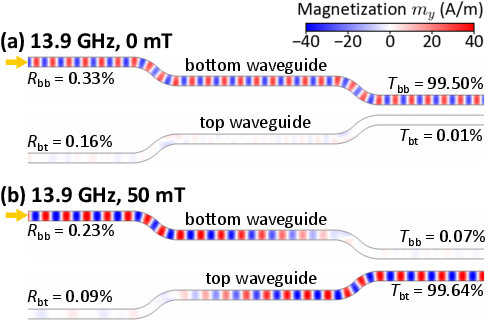}
    \caption{Propagation of the spin wave through the directional coupler at 13.9~GHz (a) in the zero field and (b) in 50~mT field. Each plot shows the bottom waveguide (shown in blue in Fig.~\ref{fig:design}(a)) at the top and the top waveguide (shown in orange in Fig.~\ref{fig:design}(a)) at the bottom. In each case, the spin wave is excited on the left side of the bottom waveguide, as indicated by the yellow arrow. \label{fig:coupler}}
\end{figure}

The functionality of the directional coupler is shown in Fig.~\ref{fig:coupler}. Firstly, without an external magnetic field [Fig.~\ref{fig:coupler}(a)], there is no coupling between the bottom and top waveguide. As a result, 99.50\% of the excited wave remains in the bottom waveguide ($T_{\rm bb}$), with only 0.01\% transmitted to the top waveguide ($T_{\rm bt}$). However, as much as 0.49\% is reflected and goes to both the top ($R_{\rm bt}$) and bottom ($R_{\rm bb}$) waveguides. When the external magnetic field is increased up to 50~mT [Fig.~\ref{fig:coupler}(b)], the coupling between the waveguides enables the transfer of the spin wave between them. As a result, 99.64\% of the wave goes to the top waveguide, with only 0.07\% remaining in the bottom one. In this case, the reflectivity drops down to 0.32\%. Therefore, these results confirm the possibility of controlling the operation of the directional coupler by the external magnetic field.

\begin{figure}
    \includegraphics[width=\linewidth]{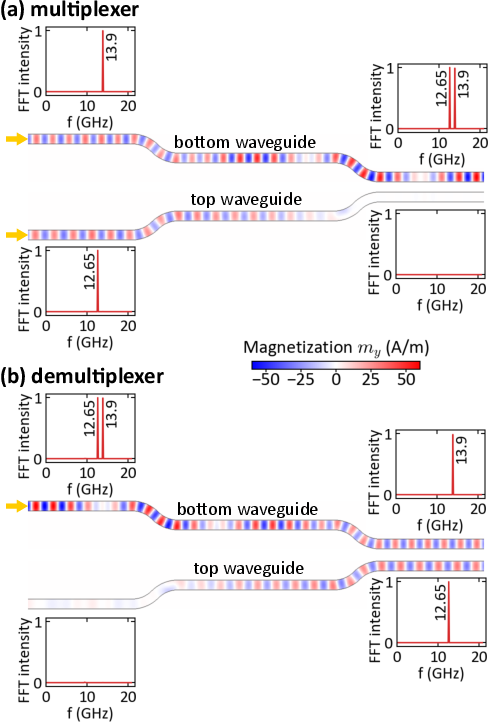}
    \caption{The directional coupler working as (a) a multiplexer and (b) a demultiplexer. In (a), the spin wave at 13.9~GHz is excited in the bottom waveguide, and the spin wave at 12.65~GHz is excited in the top waveguide. In (b), both frequencies are excited in the bottom waveguide. In both cases, the external magnetic field is set to zero. In each plot, the bottom waveguide is shown at the top and the top waveguide is shown at the bottom. The spin-wave inputs are marked with yellow arrows.
    \label{fig:multiplexer}}
\end{figure}

As mentioned above, this directional coupler can also perform the multiplexing and demultiplexing of the signal of two frequencies---12.65~GHz and 13.9~GHz at zero external field. As shown in Fig.~\ref{fig:coupler}(a), the spin wave excited at the bottom waveguide at the frequency of 13.9~GHz remains in the bottom waveguide. At the same time, the spin wave of 12.65~GHz frequency is transferred from the bottom to the top waveguide with an efficiency of 99.94\%. These multiplexing and demultiplexing functionalities are shown in Fig.~\ref{fig:multiplexer}. In the case of the signal multiplexing shown in Fig.~\ref{fig:multiplexer}(a), the spin wave at 13.9~GHz is excited in the bottom waveguide, and the spin wave at 12.65~GHz is excited in the top waveguide. As a result, after passing through the coupler region, both waves propagate in the bottom waveguide, while the top waveguide remains with almost no spin-wave intensity. Similarly, demultiplexing is shown in Fig.~\ref{fig:multiplexer}(b). This time both frequencies are excited in the bottom waveguide. As it passes through the coupler, the 13.9~GHz spin wave remains in the bottom waveguide, while the 12.65~GHz spin wave transfers into the top waveguide. In both cases, i.e., multiplexing and demultiplexing, the efficiency of the coupler is very high as the transmittance is 99.94\% for 12.65~GHz and 99.50\% for 13.9~GHz.


To sum up, we have shown that the system of vertically coupled nanoscale waveguides opens new possibilities for designing spin-wave devices. In contrast to systems of planar waveguides, vertically coupled conduits have a critical wavevector $k_{\rm c}$, at which the coupling is completely suppressed. Moreover, waveguides with dimensions below 100~nm have a wide range of frequencies in which only the fundamental mode exists. These properties can be used to design a spin-wave vertical directional coupler with a length below \SI{1}{\micro\meter}, which can be controlled by the external magnetic field. The results of micromagnetic simulations show that the efficiency of the coupler operating at the frequency of 13.9~GHz is as high as 99.5\%. Furthermore, this device can also perform the multiplexing and demultiplexing of the two-frequency signals with the same efficiency at zero external field. The demonstrated directional coupler exhibits the fundamental functionalities essential for the magnonic circuits, demonstrating the advantages of 3D systems, and paving the way for efficient spin-wave computing.

\section*{Supplementary Material}
The Supplementary Material contains the details of the numerical methods used to obtain the results, the analysis of the transmission time parameter, and the static magnetization configuration of the directional coupler.

\section*{Author contributions}
Krzysztof Szulc: Conceptualization (equal); Data curation (lead); Formal analysis (lead); Funding acquisition (lead); Investigation (lead); Methodology (lead); Project administration (lead); Software (lead); Visualization (lead); Writing -- original draft (lead); Writing -- review \& editing (equal).

Maciej Krawczyk: Conceptualization (equal); Supervision (lead); Writing -- review \& editing (equal).

\section*{Data availability}
The data that support the findings of this study are openly available in Zenodo at https://doi.org/10.5281/zenodo.14620996


\begin{acknowledgments}
K. Szulc acknowledges the financial support from National Science Centre, Poland, grant no. UMO-2021/41/N/ST3/04478.
\end{acknowledgments}

%

\end{document}



\title{SUPPLEMENTARY MATERIAL\\Magnetic field--controlled nanoscale spin-wave vertical directional coupler} 



\author{Krzysztof Szulc}
\email{szulc@ifmpan.poznan.pl}
\affiliation{Institute of Molecular Physics, Polish Academy of Sciences, M. Smoluchowskiego 17, 60-179, Poznań, Poland}
\affiliation{Institute of Spintronics and Quantum Information, Faculty of Physics and Astronomy, Adam Mickiewicz University, Uniwersytetu Poznańskiego 2, 61-614, Poznań, Poland}

\author{Maciej Krawczyk}
\affiliation{Institute of Spintronics and Quantum Information, Faculty of Physics and Astronomy, Adam Mickiewicz University, Uniwersytetu Poznańskiego 2, 61-614, Poznań, Poland}


\date{\today}

\pacs{}

\maketitle 


\section{Numerical methods}

The first part of the numerical study was related to the caluclation of the dispersion relation of the system of two interacting, identical waveguides. This was done using COMSOL Multiphysics, a finite-element method solver. We used our own implementation of the linearized Landau--Lifshitz--Gilbert equation coupled to the equation for the magnetic scalar potential in the two-dimensional geometry.\cite{vanatka2021spin} The linearization of the system was performed assuming the uniform magnetization of the waveguides along their main $x$-axis, i.e., $\mathbf{m} = (M_{\rm s},m_y,m_z)$ where the saturation magnetization $M_{\rm s} \gg m_y,m_z$. After the linearization, the Landau--Lifshitz--Gilbert equation takes a form
\begin{equation} \label{eq:ll}
\begin{split}
    \partial_t m_y 
    = 
    \gamma\mu_0 \left( H_0 m_z - 
    \frac{2A_{\rm ex}}{\mu_0 M_{\rm s}^2}
    \left(k^2+\partial_y^2+\partial_z^2\right) m_z + 
    M_{\rm s}\partial_z \varphi\right), \\
    \partial_t m_z 
    = 
    -\gamma\mu_0 \left (H_0 m_y - 
    \frac{2A_{\rm ex}}{\mu_0 M_{\rm s}^2}
    \left(k^2+\partial_y^2+\partial_z^2\right) m_y + 
    M_{\rm s}\partial_y \varphi \right),
\end{split}
\end{equation}
while the equation for the magnetic scalar potential $\varphi$ is
\begin{equation}
    \left(k^2+\partial_y^2+\partial_z^2\right) \varphi
    =
    \partial_y m_y + \partial_z m_z,
\end{equation}
where $\partial_a$ is the partial derivative with respect to the variable $a$, $\gamma$ is the gyromagnetic ratio, $H_0$ is the external magnetic field, $A_{\rm ex}$ is the exchange constant, and $k$ is the wavevector.

The waveguides are placed in the center of the $10\times\SI{10}{\micro\meter}$ vacuum chamber for proper calculation of the dynamic stray field. For the discretization of the system, we used the triangular mesh with an element size of 1.5~nm in the waveguides and an element growth rate of 1.3 in the vacuum. The dispersion is calculated using the eigenfrequency solver, where wavevector $k$ is a sweeping parameter.

The second part of the study was to analyze the propagation of the spin waves through the directional coupler. For this purpose, we performed the micromagnetic simulations in Boris Computational Spintronics.\cite{boris} We assumed a computational cell size of $2\times2\times5$~nm. The length of the computed system was $\SI{10}{\micro\meter}$. At the ends of each waveguide, the absorbing conditions were applied, which consists of the exponentially increasing damping constant, starting 1~$\SI{}{\micro\meter}$ from the boundary and reaching the value of 0.25 at the boundary. Each system was first relaxed for a given value of the external magnetic field using the Steepest Descent method with the $m\times h=10^{-9}$ as a stop parameter. Then, the spin waves were excited by an ac magnetic field, whose profile along the waveguide was limited by the Gaussian distribution with a standard deviation of 10~nm. For this purpose, we used the Runge--Kutta RK4 method. In the calculation of the transmittance as a function of frequency, the antenna was exciting the \textit{sinc} signal with a cut-off frequency of 20~GHz, the magnetic field amplitude of $\SI{10}{\micro\tesla}$, the time of 5~ns, and the center of the \textit{sinc} signal at the 2.5~ns. In the simulations of the directional coupler functionalities, the antenna was continuously exciting the sine signal with the amplitude of $\SI{10}{\micro\tesla}$.

\begin{figure}[!b]
    \includegraphics{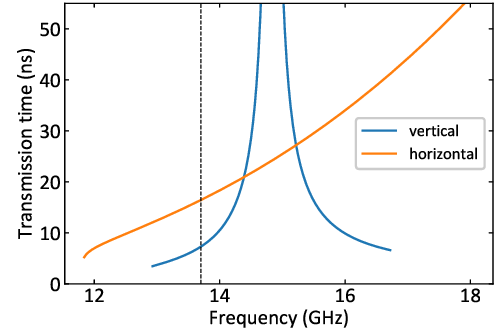}
    \caption{Transmission time as a function of frequency for the horizontally aligned (orange line) and vertically aligned (blue line) waveguides in the external magnetic field of 50~mT. Vertical dashed black line indicates the nominal directional coupler frequency of 13.7~GHz.
    \label{fig:transtime}}
\end{figure}

\begin{figure*}[!t]
    \includegraphics{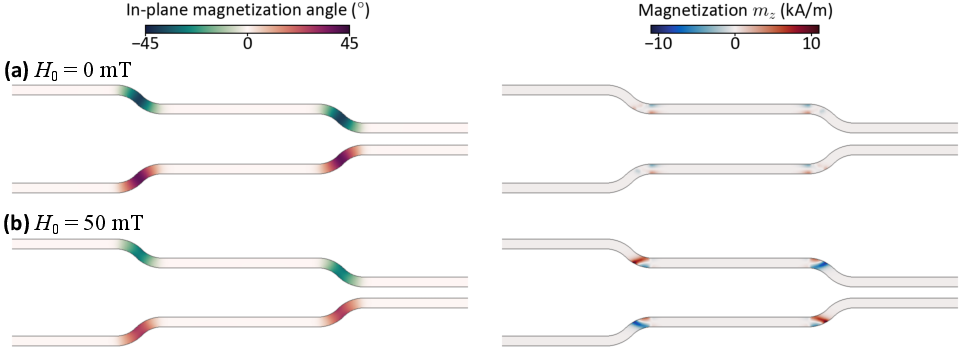}
    \caption{Static magnetization configuration of the directional coupler shown in Fig.~3 in the main text in the external magnetic field of (a) 0~mT and (b) 50~mT. On the left side, the in-plane magnetization angle is shown. The angle of $0^{\circ}$ describes the magnetization parallel to the external magnetic field, i.e., directed to the right. On the right side, the out-of-plane $m_z$ component of the static magnetization is shown.
    \label{fig:static}}
\end{figure*}

\section{Transmission time}

The coupling length does not give the complete information about the transfer of the spin wave between the waveguides since the spin-wave group velocity $v_{\rm g}$ depends on the frequency. It is important to study another parameter which is transmission time $t_{\rm t}$, which is defined as
\begin{equation}
    t_{\rm t} = \frac{l_{\rm c}}{v_{\rm g}},
\end{equation}
where $l_{\rm c}$ is a coupling length. $t_{\rm t}$ describes the time required to transfer the wave from one waveguide to another. This parameter as a function of frequency is shown in Fig.~\ref{fig:transtime} for both the horizontally aligned and vertically aligned waveguides in the external magnetic field of 50~mT. It is calculated only for the fundamental modes in the two-mode range. The transmission time is lower for the vertically coupled waveguides at the ends of the two-mode range than for any frequency for the planar waveguides. For the nominal frequency of the directional coupler calculated from the dispersion relation -- 13.7~GHz (marked with vertical dashed black line) -- the transmission time is equal to 7.27~ns, and it is slightly larger than the minimum value for planar waveguides, which is 5.21~ns. However, as it was mentioned in the main part of the manuscript, the minimum length of the horizontal directional coupler will be equal to two coupling lengths, therefore its operation time is larger than the one of the vertical directional coupler shown in this paper. One can also notice that the vertical directional coupler based on the same principle as the horizontal coupler can have smaller operation time, despite being larger, as it can be estimated from Fig.~2 in the main text.

\section{Static magnetization configuration of directional coupler}

Figure~\ref{fig:static} shows the static magnetization configuration of the directional coupler in the external magnetic field of 0 (a) and 50~mT (b). We present separately the in-plane and out-of-plane magnetization components. At the bends of the directional coupler, the magnetization also bends to be aligned with the bend. At the center of the bend, the maximum deviation angle is 39.9 degrees for zero field, which is only slightly smaller than the maximum angle of the bend (45 degrees). With the increase of the external field, the deviation decreases and for 50~mT it is 29.5 degrees. Moreover, a small out-of-plane component of the static magnetization is present in the region where the two waveguides meet. At zero magnetic field, the maximum value of $m_z$ is 3.6~kA/m, while for 50~mT it increases to 11.3~kA/m. As one can notice, this is only 1.3\% of the saturation magnetization, so the deviation is rather small. However, it can still play an important role in the spin-wave reflection at the bends.

%